\documentclass[prl,twocolumn,noshowpacs,noshowkeys,preprintnumbers,floatfix,
  nofootinbib,superscriptaddress]{revtex4-1}
\usepackage{textcomp} %
\usepackage{amsfonts} 
\usepackage{amssymb} 
\usepackage{amsmath} 
\usepackage{subfigure} 
\usepackage{graphicx} 
\usepackage{array} 
\usepackage{dcolumn} 
\usepackage{bm} 
\usepackage{latexsym} 
\usepackage{longtable} 
\usepackage{hyperref} 
\usepackage{multirow}
\graphicspath{{./figs/}}

\newcommand{\MeV}{\mathop{\rm MeV}\nolimits}
\newcommand{\GeV}{\mathop{\rm GeV}\nolimits}

\begin{document}


\title{Neutral kaon mixing from new physics: matrix elements
  in $N_f=2+1$ QCD }
\author{Taegil Bae}
\affiliation{
  Korea Institute of Science and Technology Information,
  Daejeon, 305-806, South Korea 
}
\author{Yong-Chull Jang}
\affiliation{
  Lattice Gauge Theory Research Center, FPRD, and CTP, \\
  Department of Physics and Astronomy,
  Seoul National University, Seoul, 151-747, South Korea
}
\author{Hwancheol Jeong}
\affiliation{
  Lattice Gauge Theory Research Center, FPRD, and CTP, \\
  Department of Physics and Astronomy,
  Seoul National University, Seoul, 151-747, South Korea
}
\author{Chulwoo Jung}
%
%
%
\affiliation{
  Physics Department, Brookhaven National Laboratory,
  Upton, NY11973, USA
}
\author{Hyung-Jin Kim}
\affiliation{
  Physics Department, Brookhaven National Laboratory,
  Upton, NY11973, USA
}
\author{Jangho Kim}
\affiliation{
  Lattice Gauge Theory Research Center, FPRD, and CTP, \\
  Department of Physics and Astronomy,
  Seoul National University, Seoul, 151-747, South Korea
}
\author{Jongjeong Kim}
\affiliation{
  Lattice Gauge Theory Research Center, FPRD, and CTP, \\
  Department of Physics and Astronomy,
  Seoul National University, Seoul, 151-747, South Korea
}
\author{Kwangwoo Kim}
\affiliation{
  Lattice Gauge Theory Research Center, FPRD, and CTP, \\
  Department of Physics and Astronomy,
  Seoul National University, Seoul, 151-747, South Korea
}
\author{Sunghee Kim}
\affiliation{
  Lattice Gauge Theory Research Center, FPRD, and CTP, \\
  Department of Physics and Astronomy,
  Seoul National University, Seoul, 151-747, South Korea
}
\author{Weonjong Lee}
%
%
%
%
\affiliation{
  Lattice Gauge Theory Research Center, FPRD, and CTP, \\
  Department of Physics and Astronomy,
  Seoul National University, Seoul, 151-747, South Korea
}
\author{Jaehoon Leem}
\affiliation{
  Lattice Gauge Theory Research Center, FPRD, and CTP, \\
  Department of Physics and Astronomy,
  Seoul National University, Seoul, 151-747, South Korea
}
\author{Stephen R. Sharpe}
%
%
%
\affiliation{
  Physics Department,
  University of Washington,
  Seattle, WA 98195-1560, USA
}
\author{Boram Yoon}
\affiliation{
  Lattice Gauge Theory Research Center, FPRD, and CTP, \\
  Department of Physics and Astronomy,
  Seoul National University, Seoul, 151-747, South Korea
}
%
%
%
%
%
%
\collaboration{SWME Collaboration}
\date{\today}
\begin{abstract}
We present results for matrix elements
of $\Delta S=2$ four-fermion operators arising 
generically in models of new physics. 
These are needed to constrain such models
using the measured values of $\varepsilon_K$ and $\Delta M_K$.
We use lattice QCD with $2+1$ flavors of improved
staggered fermions on lattices generated by the MILC collaboration.
We extrapolate to the continuum
from three lattice spacings ranging down to $a\approx 0.045\;$fm.
Total errors are $\sim 5-6\%$,
arising primarily from our use of one-loop matching between lattice
and continuum operators.
For two of the matrix elements, our results disagree significantly
from those obtained using different fermion discretizations.
\end{abstract}
\maketitle
%

Processes that are highly suppressed in the standard model (SM)
provide a window into beyond-the-standard model (BSM) physics
complementary to that from direct searches.
Indeed, in a given model, such processes 
can set lower limits on the scale of BSM physics
which are beyond the reach of present accelerators
(see, e.g., Refs.~\cite{Mescia:2012fg,Buras:2013td,Kersten:2012ed}).
Here we focus on kaon mixing,
which provides among the most powerful constraints.
For both the CP-conserving mass difference $\Delta M_K$
and the CP-violating part parametrized by $\varepsilon_K$,
the sum of contributions from SM and BSM physics
must add to the observed values.
BSM contributions involve heavy particles and are thus
short-distance dominated.
In order to determine the contributions in a given model,
one needs to calculate, in QCD, the matrix elements of local
$\Delta S=2$ four fermion operators.
A generic BSM model introduces four operators (listed below)
having Dirac structures different from those that arise in the SM.
Lattice QCD is the only available quantitative tool for calculating
such matrix elements from first principles, and we present here 
such a calculation.

In a generic BSM model (be it 
supersymmetry, extra dimensions, little Higgs, etc.)
$\Delta S=2$ processes involve loops of heavy particles,
that, when integrated out, lead to the effective Hamiltonian
\begin{align}
 H^{BSM}_{\Delta S=2} &= \sum_i C_i(\mu) Q_i\,, \\
 {Q}_{1} &=
 [\bar{s}^a \gamma_\mu (1-\gamma_5) d^a]
 [\bar{s}^b \gamma_\mu (1-\gamma_5) d^b]\,,   \\
 {Q}_{2} &=
 [\bar{s}^a (1-\gamma_5) d^a] [\bar{s}^b (1-\gamma_5) d^b]\,,   \\
 {Q}_{3} &=
 [\bar{s}^a \sigma_{\mu\nu}(1-\gamma_5) d^a]
 [\bar{s}^b \sigma_{\mu\nu} (1-\gamma_5) d^b]\,,   \\
 {Q}_{4} &=
 [\bar{s}^a (1-\gamma_5) d^a] [\bar{s}^b (1+\gamma_5) d^b]\,,  \\
 {Q}_{5} &=
 [\bar{s}^a \gamma_\mu (1-\gamma_5) d^a]
  [\bar{s}^b \gamma_\mu (1+\gamma_5) d^b]
\,.
\end{align}
Here $C_i$ are Wilson coefficients that can be calculated in
a given theory (with $\mu$ the renormalization scale),
$a,b$ are color indices,
$\sigma_{\mu\nu}= [\gamma_\mu,\gamma_\nu]/2$,
and repeated indices are summed.
The local $\Delta S=2$ operators $Q_i$, are
in the chiral basis of Ref.~\cite{Buras:2000if}.\footnote{%
Our normalization differs from Ref.~\cite{Buras:2000if},
but this has no impact on the associated $B$-parameters.}
%
%
%
%
%
%
%
%
$Q_1$ is the ``left-left'' operator arising in the SM
contribution to $\varepsilon_K$, while
$Q_{2-5}$ are the BSM operators.
In previous lattice calculations, a different basis
(the ``SUSY basis'') for the operators has been used.
We prefer the chiral basis as it has been used
to calculate the two-loop anomalous dimensions
which we use to run the results between different 
renormalization scales~\cite{Buras:2000if}.
Results can be easily converted to the SUSY basis at
the end---see below.

In our lattice set-up (described below) we obtain directly
ratios of matrix elements. Specifically, we calculate the
following $B$-parameters for the BSM operators: 
%
\begin{align}
\label{eq:def-B_i}
  B_i(\mu) = 
\frac{\langle \overline{K}_0 \vert Q_i (\mu) \vert K_0 \rangle} 
{N_i \langle \overline{K}_0 \vert\overline{s}\gamma_5 d(\mu)\vert 0 \rangle 
    \langle 0 \vert \bar{s} \gamma_5 d (\mu) \vert K_0 \rangle}
\\
  (N_2,\ N_3,\ N_4,\ N_5) = ( 5/3, \ 4, \ -2, \ 4/3 )
\,.
\end{align}
The operators in both numerator and denominator are renormalized
in the $\overline{\rm MS}$ scheme using naive dimensional
regularization for $\gamma_5$ and other conventions described
in Ref.~\cite{Buras:2000if}.
The matrix elements of $Q_i$ can be obtained from the
$B$-parameters since the denominators can be expressed in terms
of known quantities ($M_K$, $f_K$, $m_s+m_d$).\footnote{%
In other calculations of the BSM 
matrix elements~\cite{Boyle:2012qb,Bertone:2012cu},
different ratios have been used so as to avoid the need to use quark masses.
However, quark masses are now quite well determined~\cite{Beringer:1900zz},
with the $\sim 6\%$ error in $(m_s+m_d)^2$ being comparable
to those we obtain for the $B_i$.}
We also calculate $B_K$.

There have been two previous calculations of the BSM matrix elements
using dynamical quarks, one using domain-wall fermions~\cite{Boyle:2012qb}
the other twisted-mass fermions~\cite{Bertone:2012cu}.\footnote{%
There have also been earlier quenched calculations, which established
the basic methodology~\cite{Conti:1998ys,Donini:1999nn,Babich:2006bh}.
}
The former uses the physical complement
of $2+1$ light sea quarks,
but has results only at a single lattice spacing,
The latter uses $2$ dynamical flavors,
with the strange quark being quenched.
Our calculation uses 
$2+1$ flavors of dynamical staggered quarks, with multiple lattice
spacings.
It is thus the first to control all sources of error.
We have previously used similar methodology 
to calculate $B_K$~\cite{Bae:2011ff},
with results that are consistent with 
those from other types of fermion.

An important advantage of staggered fermions is that they are
computationally cheap, 
allowing calculations at multiple lattice spacings, quark
masses and volumes.\footnote{%
Staggered fermions also preserve part of the continuum
chiral symmetry, although this is less
important for the BSM matrix elements than for $B_K$, as the
former are not constrained by chiral symmetry to vanish
in the $SU(3)$ chiral limit.}
Their main disadvantage is that each flavor
comes in 4 copies (``tastes''), with the associated $SU(4)$ symmetry
broken at non-zero lattice spacing $a$. Removing the extra tastes requires
rooting the fermion determinant, but we assume that the
artifacts this introduces
vanish in the continuum limit $a\to0$. 

We use ensembles generated
with the improved ``asqtad'' staggered action
by the MILC collaboration~\cite{Bazavov:2009bb}.
Those used here are listed in Table~\ref{tab:milc-lat},
with $m_\ell$ the average up/down sea-quark mass 
and $m_s$ the strange sea-quark mass
(which lies close to the physical value on all ensembles).
For valence quarks, we use HYP-smeared staggered 
fermions~\cite{Hasenfratz:2001hp}.
These are known to substantially
reduce both artifacts due to taste symmetry breaking
and perturbative corrections to matching 
factors~\cite{Hasenfratz:2001hp,Bae:2008qe}.

%
\begin{table}[tbp]
\caption{MILC ensembles used in this work, with
 ``ens'' the number of gauge configurations and
``meas'' the number of measurements per configuration.
  ID identifies the ensemble, with 
F=fine, S=superfine and U=ultrafine.
  \label{tab:milc-lat}}
\begin{ruledtabular}
\begin{tabular}{c  c  c  c  l }
$a$ (fm) & $am_\ell/am_s$ & \ \ size & ens $\times$ meas  & ID \\
\hline
0.09  & 0.0062/0.031 & $28^3 \times 96$  & $995 \times 9$ & F1 \\
0.09  & 0.0093/0.031 & $28^3 \times 96$  & $949 \times 9$ & F2 \\
0.09  & 0.0031/0.031 & $40^3 \times 96$  & $959 \times 9$ & F3 \\
0.09  & 0.0124/0.031 & $28^3 \times 96$  & $1995\times 9$ & F4 \\
0.09  & 0.00465/0.031 & $32^3 \times 96$  & $651\times 9$ & F5 \\
\hline
0.06  & 0.0036/0.018 & $48^3 \times 144$ & $749 \times 9$ & S1 \\
0.06  & 0.0072/0.018 & $48^3 \times 144$ & $593 \times 9$ & S2 \\
0.06  & 0.0025/0.018 & $56^3 \times 144$ & $799 \times 9$ & S3 \\
0.06  & 0.0054/0.018 & $48^3 \times 144$ & $582 \times 9$ & S4 \\
\hline
0.045 & 0.0028/0.014 & $64^3 \times 192$ & $747 \times 1$ & U1 \\
\end{tabular}
\end{ruledtabular}
\end{table}

Our mixed action set-up is identical to that we used previously to calculate 
$B_K$, and the detailed
lattice methodology is also very similar~\cite{Bae:2010ki,Bae:2011ff}.
On each lattice, we place two wall sources separated by an
fixed interval $\Delta t$. 
These create 
kaons having taste $\xi_5$ and zero spatial momenta.
Lattice versions of the BSM operators are placed between 
the two wall sources, as are the pseudoscalar operators needed
for the denominators of the $B_i$ [see Eq.~(\ref{eq:def-B_i})].
We choose $\Delta t$ such that the contamination from excited
states and from kaons ``propagating around the world''
can be ignored. We use the same values as in our $B_K$ calculation,
the justification for which has been discussed in 
Ref.~\cite{Bae:2010ki}.
Multiple measurements with random time translations are carried out
on each lattice (see Table~\ref{tab:milc-lat}).
After averaging over configurations, we form the ratios needed for
the $B_i$. The overlap of the wall sources with the kaon states
cancels in these ratios, as well as some of the statistical error.
Away from the sources, these ratios should be independent of
$t$. We find this to be the case within errors,
and we fit them to a constant over a central ``plateau'' region.
%

Our lattice operators are matched to those in the continuum
scheme of Ref.~\cite{Buras:2000if} using 1-loop,
mean-field improved perturbation theory.
Previous one-loop matching calculations matched the lattice operators
to those in a different
continuum scheme~\cite{Kim:2011pz},
but we have now determined the matching between the
two continuum schemes~\cite{bsm-toolkit}.
At each lattice spacing, we match to the continuum scheme at scale
$\mu=1/a$.
The one-loop corrections are typically 10-20\%,
with the largest being 30\%. This is in line with the expectation
that the coefficient of $\alpha$ should be of ${\cal O}(1)$ or smaller.

We use 
a partially quenched set-up with 
ten different valence quark masses:
$am_{x,y} = am_s \times (n/10)$ and $n=1,2,\ldots,10$. 
Here $x$ and $y$ refer to valence $d$ and $s$ quarks, respectively.
Our lightest valence pions have $M_{x\bar x}\approx 200\;$MeV.
For our valence kaons, 
we use the lightest 4 values of $am_x$ and the heaviest 3 of $am_y$.
These combinations satisfy
$m_x \ll m_y\sim m_s^\textrm{phys}$, 
so that our result lie in the regime in which heavy-kaon
SU(2) chiral perturbation theory (ChPT) is 
applicable~\cite{Roessl:1999iu,Allton:2008pn}.

For our chiral extrapolations, we use $B_K$ (whose behavior is
well understood from prior work) as well as the four ``golden'' combinations
\begin{align}
G_{23} = \frac{B_2}{B_3}, \;
G_{45} = \frac{B_4}{B_5}, \;
G_{24} = B_2 B_4, \;
G_{21} = \frac{B_2}{B_K}.
\label{eq:golden}
\end{align}
These have no chiral logarithms 
at next-to-leading order (NLO)~\cite{Bailey:2012wb}, 
and thus have simpler chiral extrapolations 
than the $B_i$ and reduced sensitivity to taste-breaking lattice artifacts.
At the end, we invert these relations to determine the $B_i$.
%
We have checked that extrapolating the $B_i$ directly leads to
compatible results~\cite{BSMinprep}.

\begin{figure}[t!]
  \includegraphics[width=20pc]{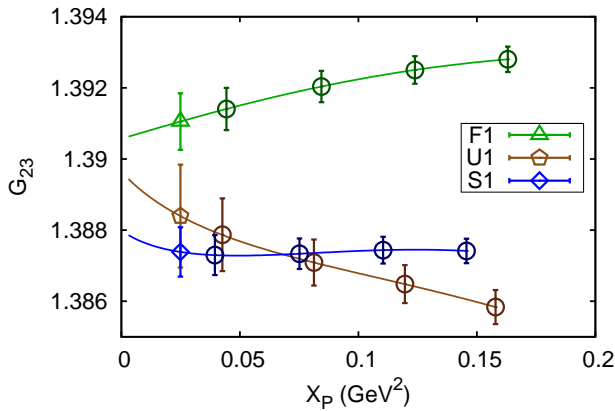}
  \caption{ $G_{23}({\rm NDR},1/a\;{\rm GeV})$ vs. $X_P$ on the F1, S1,
    and U1 ensembles. Data point are shown by circles, while
    the triangle, pentagon, and diamond show the extrapolated results.}
  \label{fig:g23-X-fit}
\end{figure}

We extrapolate to physical quark masses and $a=0$ in three steps.
The first two are done on each ensemble separately:
we extrapolate $m_x$ to $m_d^{\rm phys}$ 
(``X-fit'') and then $m_y$ to $m_s^{\rm phys}$ (``Y-fit'').
For $B_K$ the fitting is 
as described in Refs.~\cite{Bae:2010ki,Bae:2011ff}.
For the $G_i$ we fit to
\begin{equation}
c_1 + c_2 X + c_3 X^2 + c_4 X^2 \ln^2 X 
+ c_5 X^2 \ln X + c_6 X^3 \,, 
\end{equation}
where $X \equiv X_P / \Lambda_\chi^2$, with 
$X_P = M_{x\bar x}^2$ and $\Lambda_\chi = 1\;$GeV.
This incorporates the absence of the NLO logarithm.
The NNLO chiral logarithms are not known, so we
use the generic form of such terms.
We also include a single analytic NNNLO term,
which is required for good fits.
Our X-fits include the full correlation matrix, 
and constrain the coefficients
$c_{3-6}$ with Bayesian priors: $c_i = 0 \pm 1$.
Examples of X-fits for $G_{23}$ are shown in Fig.~\ref{fig:g23-X-fit}.
We see that the chiral extrapolations are short and the dependence mild.
This holds also for the other $G_i$.
Systematic errors are estimated by doubling the widths of the Bayesian priors,
and also by comparing with fits using the eigenmode 
shift method~\cite{Jang:2011fp}. These two estimates are then combined
in quadrature.

The $m_y$ dependence is close to linear, so
we use a linear extrapolation for our central value and
a quadratic fit to estimate a systematic error.
After these Y-fits, we evolve results from all lattices to
a common renormalization scale, either $\mu = 2$ or $3\; \GeV$,
using the two-loop anomalous dimensions given in
Ref.~\cite{Buras:2000if}.
It turns out that the standard form of solution is singular,
but this can be removed by the analytic continuation method
proposed in Ref.~\cite{Adams:2007tk}.
Details will be given in Ref.~\cite{bsm-toolkit}.

\begin{figure}[tb!]
  \includegraphics[width=20pc]{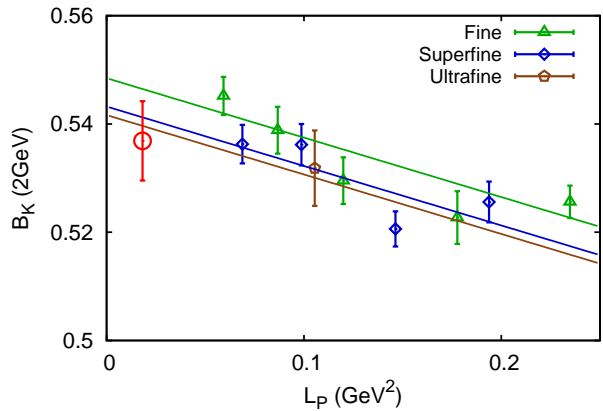}
  \caption{Chiral-continuum extrapolation
of $B_K({\rm NDR},2\;{\rm GeV})$.
    The red circle shows the extrapolated result. 
  }
  \label{fig:bk-lp}
\end{figure}

\begin{figure}[tb!]
  \includegraphics[width=20pc]{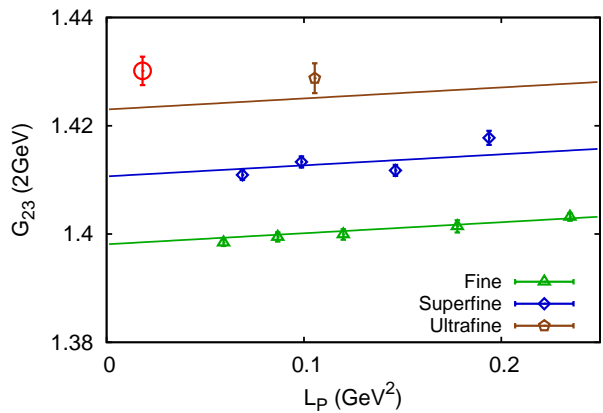}
  \caption{ As for Fig.~\protect\ref{fig:bk-lp} but for $G_{23}$. 
}
  \label{fig:g23-lp}
\end{figure}

In our final extrapolation we simultaneously
extrapolate sea-quark masses to their physical values
and $a\to 0$. 
As for $B_K$, 
we use only the three finest lattice spacings.
For our central values we fit to
\begin{equation}
f_1=d_1 + d_2 (a \Lambda_Q)^2 + d_3 {L_P}/{\Lambda_\chi^2}
+ d_4 {S_P}/{\Lambda_\chi^2}\,,
\label{eq:fitf1}
\end{equation}
with $L_P$ ($S_P$) the squared masses of the taste
$\xi_5$ pseudoscalars composed of light (strange) sea quarks.
We set $\Lambda_Q = 0.3 \GeV$ 
and expect $d_2\sim {\cal O}(1)$. In fact, for the $G_i$
we find $|d_2|\sim 2-7$, indicating enhanced discretization errors.
We also find that
$d_3\ne d_4$, indicating substantial SU(3) breaking,
although both coefficients are of the expected size $|d_{3,4}|\ll 1$.
Examples of fits are shown in Figs.~\ref{fig:bk-lp}
and \ref{fig:g23-lp}, for $B_K$ and $G_{23}$ respectively.
These fits
 have $\chi^2/\text{dof}=1.9$ and $2.7$ respectively. 
Those for the other $G_i$ are slightly better, with
$\chi^2/\text{dof}=1.6-1.7$.   

\begin{table}[tbp]
\caption{Final results for the $B_i$ and $G_i$
at two renormalization scales.
Errors are respectively statistical and systematic.}
  \label{tab:g_and_b-par}
\begin{ruledtabular}
\begin{tabular}{ l c c c c }
      & $\mu=2\GeV$ & $\mu=3\GeV$ \\
\hline
$B_K$ & 0.537 (7)(24)  & 0.519 (7)(23) \\
$B_2$ & 0.620 (4)(31)  & 0.549 (3)(28) \\
$B_3$ & 0.433 (3)(19)  & 0.390 (2)(17) \\
$B_4$ & 1.081 (6)(48)  & 1.033 (6)(46) \\
$B_5$ & 0.853 (6)(49)  & 0.855 (6)(43) \\
\hline
$G_{23}$ & 1.430 (3)(63)  & 1.405 (2)(62) \\
$G_{45}$ & 1.267 (2)(72)  & 1.209 (1)(60) \\
$G_{24}$ & 0.670 (4)(30)  & 0.567 (4)(27) \\
$G_{21}$ & 1.155 (12)(66) & 1.058 (11)(62)\\
\end{tabular}
\end{ruledtabular}
\end{table}
%
%
%

%
%
%
\begin{table}[tbp]
\caption{Error budgets (in percent) for the $B_i(2\;\GeV)$.}
  \label{tab:err-budget}
\begin{ruledtabular}
\begin{tabular}{ l  c c c c c }
\quad source of error & $B_K$ & $B_2$ & $B_3$ & $B_4$ & $B_5$ \\
\hline
\quad statistics      & 1.37 & 0.64 & 0.63 & 0.60 & 0.66 \\
$\left\{ \begin{array}{l} \text{matching} \\ \text{cont-extrap.} \end{array} \right\}$
                      & 4.40 & 4.95 & 4.40 & 4.40 & 5.69 \\
\quad X-fit (F1)      & 0.10 & 0.10 & 0.10 & 0.12 & 0.12 \\
\quad Y-fit (F1)      & 0.62 & 0.12 & 0.19 & 0.22 & 0.16 \\
\quad finite volume   & 0.50 & 0.50 & 0.50 & 0.50 & 0.50 \\
\quad $r_1=0.3117(22)\;$fm
                      & 0.34 & 0.18 & 0.17 & 0.05 & 0.02 \\
\quad $f_\pi=132$ vs. $124\MeV$ (F1)
                      & 0.46 & 0.46 & 0.46 & 0.46 & 0.46 
\end{tabular}
\end{ruledtabular}
\end{table}

We also consider more elaborate fits, using
\begin{equation}
f_2 = f_1 + d_5 (a\Lambda_Q)^2 \alpha_s + d_6 \alpha_s^2
+ d_7 (a\Lambda_Q)^4\,,
\label{eq:fitf2}
\end{equation}
with $\alpha_s=\alpha_s(\overline{\rm MS},1/a)$.
This form includes all terms expected at NLO 
in staggered ChPT~\cite{Bailey:2012wb}
plus one NNNLO term.
We impose Bayesian constraints, $d_{2-7} = 0\pm 2$, and
find improved fits with $\chi^2/\text{dof}=0.8-1.5$.
We take the difference between $f_2$ and $f_1$ fits as the
systematic error in the chiral-continuum extrapolation.

We present our final results and error budgets
in Tables~\ref{tab:g_and_b-par} and \ref{tab:err-budget}
respectively.
Statistical errors are estimated by bootstrap and are
at the percent-level or smaller.
Of the errors not discussed above, the dominant one 
comes from our use of one-loop matching. We estimate
this as $\delta B/B=\alpha_s^2$, with $\alpha_s$
evaluated on our finest lattice~\cite{Bae:2010ki}.
Support for this estimate comes from a recent
comparison of perturbative and non-perturbative renormalization
using staggered bilinears~\cite{Lytle:2013qoa}.
Since this error is also accounted for by the
$d_6$ term when fitting to Eq.~(\ref{eq:fitf2}),
to avoid double-counting we take the largest of 
``$f_2-f_1$ error'' and the ``$\alpha^2$ error''
for our combined chiral-continuum-matching error.
Finite volume errors are estimated 
by comparing
fits using ChPT with and without finite-volume corrections.
Since the chiral logarithms in all
BSM $B$-parameters have the same relative magnitude as that in $B_K$,
we take this estimate from our earlier work on $B_K$~\cite{Bae:2011ff}.
Specifically, the largest error estimate is from the F1 ensemble
(see Table I of Ref.~\cite{Kim:2011qg}).

We close by comparing our results to those of
Ref.~\cite{Boyle:2012qb}.\footnote{%
We use this work rather than Ref.~\cite{Bertone:2012cu},
since the latter quenches the strange quark.
This choice is not quantitatively important, however, since
the results from these two works are consistent.}
As noted above, the results for $B_K$ agree.
For the BSM $B$-parameters, Ref.~\cite{Boyle:2012qb}
finds, at $\mu=3\GeV$,
$B_{i=2-5}^{\rm SUSY}=0.43(5),\; 0.75(9), \;0.69(7),\; 0.47(6)$.
Only $B_3$ differs between SUSY and chiral bases, with
$B_3^{\rm SUSY} = 
(5 B_2^{\rm chiral}-3 B_3^{\rm chiral})/2$.
Our results convert to $B_3^{\rm SUSY}(3\GeV)={0.79(3)}$.
Using this result and Table~\ref{tab:g_and_b-par}, 
we find that $B_2$ and $B_3^{\rm SUSY}$ are consistent 
with Ref.~\cite{Boyle:2012qb}
(the former only at $2\sigma$), while
$B_4$ and $B_5$ differ significantly
(by $4\sigma$ and $5\sigma$, respectively).
Our $B_4$ and $B_5$ are larger
than those of Ref.~\cite{Boyle:2012qb}
by 50\% and 80\%, respectively.

Given this difference, we have cross-checked the
components of our calculation in several ways
e.g. comparing our RG running matrices
with those in Refs.~\cite{Boyle:2012qb,Mescia:2012fg}.
Clearly, further investigation is needed to resolve the disagreement
with Refs.~\cite{Boyle:2012qb,Bertone:2012cu}.
One possibility that we are investigating is that the true
truncation errors in perturbative  matching are larger than our estimate,
which can be checked by renormalizing our operators non-perturbatively.
Another useful test would be for the other calculations to
calculate directly the golden combinations, so as to pinpoint 
the source of the disagreement.

\begin{acknowledgments}
We thank Claude Bernard for providing unpublished information.  
We thank Peter Boyle, Nicolas Garron, and Vittorio Lubicz for helpful
discussion.
W.~Lee is supported by the Creative Research Initiatives program
(2013-003454) of the NRF grant funded by the Korean government (MSIP).
C.~Jung and S.~Sharpe are supported in part by the US DOE through
contract DE-AC02-98CH10886 and grant DE-FG02-96ER40956, respectively.
Computations for this work were carried out in part on the QCDOC
computer of the USQCD Collaboration, funded by the Office of Science
of the US DOE.  W.~Lee acknowledges support from the KISTI
supercomputing center through the strategic support program
[No. KSC-2012-G3-08].
\end{acknowledgments}

%

\bibliographystyle{apsrev4-1} 
\bibliography{ref} 

\end{document}